# Training Symbol-Based Equalization for Quadrature Duobinary PDM-FTN Systems

S. Zhang, D. Chang, O. A. Dobre, O. Omomukuyo, X. Lin, and R. Venkatesan

*Abstract*—A training symbol-based equalization algorithm is proposed for polarization de-multiplexing in quadrature duobinary (QDB) modulated polarization division multiplexedfaster-than-Nyquist (FTN) coherent optical systems. The proposed algorithm is based on the least mean square algorithm, and multiple location candidates of a symbol are considered in order to make use of the training symbols with QDB modulation.Results show that an excellent convergence performance is obtained using the proposed algorithm under different polarization alignment scenarios. The optical signal-to-noise ratio required to attain a bit error rate of $2\times10^{-2}$ is reduced by 1.7 and 1.8 dB using the proposed algorithm, compared to systems using the constant modulus algorithm with differential coding for 4-ary quadrature amplitude modulation(4-QAM) and 16-QAM systems with symbol-by-symbol detection, respectively.Furthermore, comparisons with the Tomlinson-Harashima precoding-based FTN systems illustrate that QDB is preferable when 4-QAM is utilized.

## I. Introduction

RECENTLY,the use of quadrature duobinary (QDB) modulation for polarization division multiplexed (PDM) faster-than-Nyquist (FTN) systems has been proposed and investigated to increase the spectral efficiency in optical communicationnetworks [1-5]. In these studies, blind adaptive equalization techniques such as constant modulus algorithm (CMA) and multi-modulus algorithm (MMA) were applied to separate the polarization multiplexed signals. However, both the CMA and MMA may converge slowly, and suffer from the singularity problem where the equalizer recovers the same signal (or with some time shift) for both polarization outputs. Additionally, differential coding, which is employed with the CMA/MMA to handle the phase ambiguity problem,where the phase rotations of integer multiples of $\pi/2$ from the original symbol are not tracked, leads to performance penalty [6].

Another blind adaptive equalization technique, the phase-dependent decision-directed least-mean square (DD-LMS) algorithm, wasimplemented in [3], [5] for the QDB PDM 4-ary quadrature amplitude modulation (4-QAM) systems without the aid of differential coding. However, whenthe signal is highly distortedby the channel, the DD-LMS algorithm also suffers from the phase ambiguity problem, as well asthe convergence failure [7].

To avoid these problems, it is desirable to use training symbol-based algorithms. Nevertheless, this is not straightforward for the QDB modulated signals because each original training symbol has several possible candidate positions after the QDB operation, as shown in the following section.

In this letter, we propose, for the first time to our knowledge, a training symbol-based least mean square (TS-LMS) algorithm fortheQDB-PDM-FTN systems. It exhibits an excellent convergence performance and avoids the degradation ofthe system performance introduced by differential coding. A comparisonbetween the standalone DD-LMS algorithm and the proposed TS-LMS algorithmon their convergence under different polarization alignment conditions is conducted in simulation. Furthermore,thebit error rate (BER)is measured as a function of the optical signal-to-noise ratio (OSNR)of the received signal using the TS-LMS algorithm and the CMA with differential coding in 4-QAM and 16-QAM systems.In addition, wecompare the QDB-PDM-FTN systems with the PDM-FTN schemes enabled by Tomlinson-Harashima precoding [8] (THP) for both 4-QAM and 16-QAM systems in terms of BER performance and their sensitivity to channel spacing variation.

## II. Principle of the Proposed TS-LMS Algorithm

In the QDB modulated systems, each quadrature (I/Q) branch of the incoming $m$-aryQAM ($m$ = 4, 16) symbol is processed by the duobinaryoperator,in which the previous symbol delayed by one symbol period $T$ is added to the current symbol. As illustrated in Fig. 1(a),before the duobinary operation, we implement a precoder with the inverse operation, where the modulo-$\sqrt{m}$ adder is used to constrain the scope of the output. Hence, the symbol-by-symbol (SbS) detection without error propagation can be applied, and the proposed TS-LMS algorithm becomes feasible. Note that the precoder will not degrade the performance of the system [2], and a better performance can be obtained by applying the duobinary operation at the transmitter,as the processing of the transmission-added white noise is avoided.

Figs. 1 (b) and (c) respectively show the constellation maps of 4-QAM and 16-QAM signals after the QDB operation. It is worth noting that for both 4-QAM and 16-QAM signals,

This work was supported in part by the Atlantic Canada Opportunities Agency (ACOA), in part by the Research and Development Corporation (RDC).
S. Zhang, D. Chang, O. Dobre, O. Omomukuyo, X. Lin and R. Venkatesan are with the Faculty of Engineering and Applied Science, Memorial University, St. John's, NL, A1B 3X5, Canada (e-mail: szhang13@mun.com)

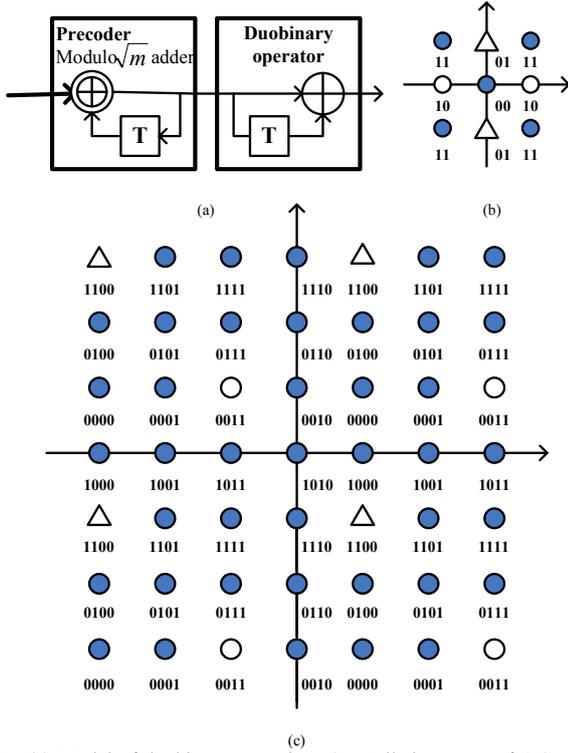

Fig. 1. (a) Model of duobinary operation; Constellation maps of 4-QAM;(b) and 16-QAM (c) signals with QDB (hollow points: selection of training symbols).

multiple points among the constellation points carry the same $\sqrt{m}$ information bits, except for the center constellation point at the origin. For example, the two hollow circle points in Fig. 2(b) carry the bits "10", and the four hollow circle points in Fig. 2(c) carry the bits "0011".In other words, the $n$-thoriginal QAM symbolis located on one of the multiple possible locations $C_p^1(n), C_p^2(n), \ldots, C_p^t(n)$, where $p$ represents either the $x$ or $y$ polarization, and $t$ = 1, 2, 4 for the symbols in the center, on the axis,and off the axis, respectively.This can be understood intuitively in the QDB operation process, where the exactlocation of anoutgoing symbol is determined by the previous random data symbols. Convergence failures of the traditional TS-LMS algorithmcan occur in the polarization multiplexed QDB systems if the multiple possible locations of training symbols are not considered.

We propose a TS-LMS algorithm that takes into account the multiple possible positions of the QDB modulated symbols. The proposed algorithm consists of two operation modes: the training mode and the tracking mode. At the beginning, the training mode makes use of the training symbols and the LMS algorithm is implemented to perform adaptive equalization. Then, it switches to the tracking mode, where the DD-LMS algorithm is used to keep track of the variation of the channel. $N_1$ training symbols are inserted at the beginning of the transmitted $m$-aryQAM symbols in order to achieve the pre-convergence at the receiver. Then, $N_2$ training symbols are inserted after every $N_b$ transmitted symbols to track thedynamic channel behaviors, such as the phase cycle slip and transient polarization state change.When updating the butterfly-type adaptive equalizer in the training mode, instead of using one fixed location, we choose the $n$-th training symbol's location $d_p(n)$ between the $t$ candidate positions by the following rule:

$$d_p(n) = \arg\min_{S_p(n)}\left(\left|S_p(n) - E'_p(n)e^{-i\varphi_p(n)}\right|\right) \quad (1)$$

where $S_p(n) = C_p^1(n), C_p^2(n), \ldots, C_p^t(n)$, $E'_p(n)$ is the $n$-thtraining symbol's corresponding received symbol in the polarization branch $p$ at the output of the butterfly-type adaptive equalizer, and $\varphi_p(n)$ is the $p$-polarization branch's carrier phase estimated by the pilot-assisted decision-aided maximum-likelihood algorithm [9].To mitigate the possible phase cycle slip and enlarge the Euclidean distance, we generate the training sequence by randomly selecting training symbols among "01" and "10" for 4-QAM and "1100" and "0011" for 16-QAM, as shown by the hollow points in Figs. 1 (b) and (c), respectively. It is noticeable that in the 16-QAM case there are other options for generating training symbols. Furthermore, the points on the in-phase (I) or quadrature (Q) axis, such as "0110" and "1001", are not suitablebecause they cannot detect and correct the 180º phase cycle slips.

### III. SIMULATION SETUP AND RESULTS

To investigate the performance of the proposed algorithm, a system model, whose schematic is depicted in Fig. 2 (a) is built using VPI TransmissionMaker. Five channels using the same scheme (QDB or THP) are simulated. The performance is assessed on the central channel. The symbol rate of each channel is 32 Gbaud. Unless otherwise mentioned, the channel spacing, $\Delta f$,is 30 GHz, and the laser linewidth is 100 kHz.The optical signal ismodulated by two ideal IQ modulators. Polarization multiplexing is performed by a polarization beam splitter (PBS) and a polarization beam combiner (PBC). The generation of the QDB and THP signals on each polarization branch are shown in Figs. 2 (b) and (c), respectively. For both schemes, $N_1$, $N_2$ and $N_b$ are chosen to be 1000, 24 and 1000.Note that the selection rules of the training symbols for THP scheme is explained in [10]. In the QDB scheme, the quadrature precoder and duobinary operator are implemented to process the symbols onI and Q branches separately. In the THP scheme, the signal is processed by the feedback filter (THP-FBF). The symbols are up-sampled by a factor of 2 and digitally shaped using the root raised-cosine (RRC) filter with a roll-off factor of 0.1 and a 3-dB bandwidth,$\Omega$. Here,$\Omega$ is fixed to 32 GHz for the QDB scheme, whereas for the THP scheme, $\Omega$ is chosen to give the best BER performance for each value of the channel spacing [8]. When $\Delta f$ is 30 GHz, the optimized $\Omega$ is 28 GHz in the THP scheme. To simplify the investigation, both chromatic dispersion and fiber nonlinear effects are neglected, and a first-order PMD emulator is employed to investigate the performance of the polarization de-multiplexing algorithms. The differential group delay (DGD) value is 50 ps.The worst polarization alignment condition is considered, where the state of polarization (SOP) ofthe input signal is offset from the principal axis of the DGD element by 45 degrees [11], unless otherwise mentioned.At the receiver, the central channel is selected by using a 0.4 nm optical band pass filter (OBPF) and then coherently detected by the combination of a local laser with zero frequency offset to the transmitter laser of the central channel, and a 90º optical hybrid. The detected signal is then processed by the receiver DSP of QDB and THP, as shown in Figs. 2 (d) and (e), respectively. For both schemes, the TS-LMS algorithm is implemented by an 11-tap butterfly-type adaptive filter. In the QDB case, the pilot-assisted decision-aided(DA) maximum-likelihood(ML) algorithm is implemented with the equalizer to perform carrier phase recovery. It is worth noting that the training stage of the equalizer can be bypassed to assess the convergence of the standalone DD-LMS. When the CMA is used for polarization de-multiplexing, the number of taps of the equalizer is 11. Pre- and post-filtering are implemented to enable the CMA for QDB modulated signals [12]. The α factor of the pre-filter [12] is optimized to be 0.7, which gives the best system performance. To solve the singularity problem, which was observed in simulation, the CMA is modified according to [13]. The carrier phase recovery is then performed using the Viterbi-and-Viterbi (V-V) algorithm for the 4-QAM and blind phase search algorithm [14] for the 16-QAM systems. Either SbS or the maximum-likelihood sequence detection (MLSD) is used to recover the bits. In the THP case, the feedforward equalizer of THP (THP-FFE) is implemented before the adaptive filter. The parameters of THP, such as the modulo size and the tap coefficients are optimized to obtain the best BER performance for the given Δf[7].

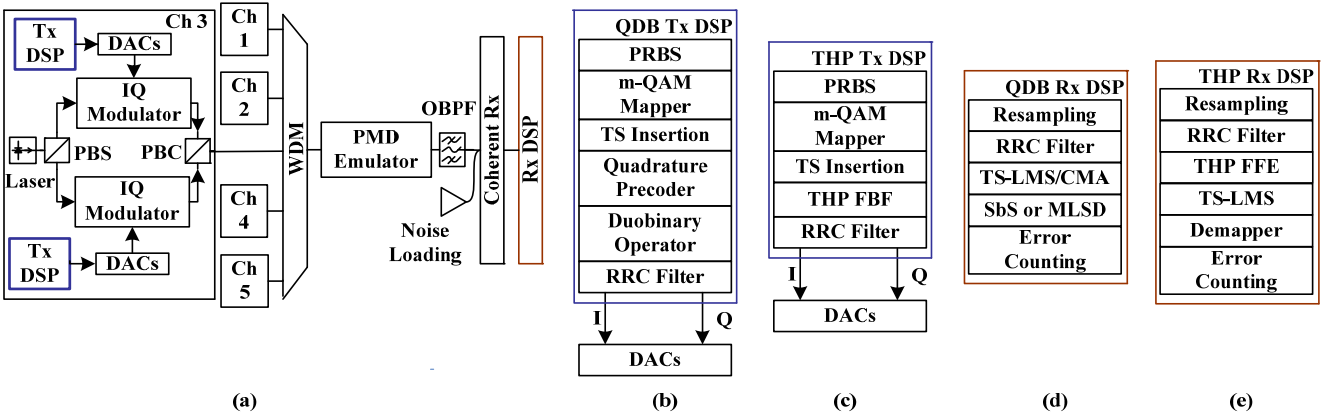

Fig. 2. (a) The schematic diagram of the FTN system. (b) QDB transmitter digital signal processing (DSP) diagram. (c) THP transmitter diagram. (d) QDB receiver DSP diagram. (e) THP receiver DSP diagram. (PBS: polarization beam splitter, PBC: polarization beam combiner, OBPF: optical bandpass filter, PRBS: pseudorandom bit sequence, RRC: root raised cosine, DAC: digital-to-analog converter, FBF: feedback filter; FFE: feed forward equalizer.

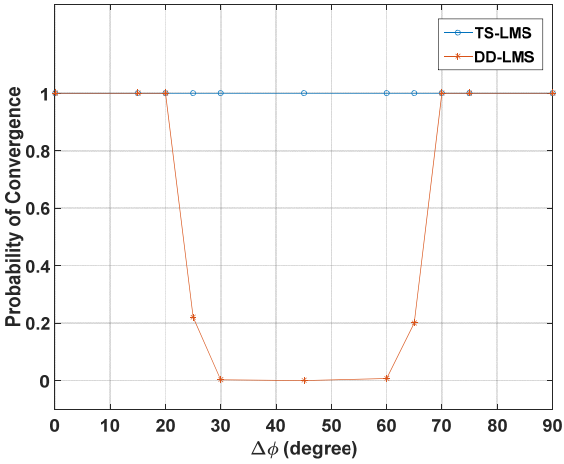

Fig. 3. Probability of convergence as a function of Δϕ.

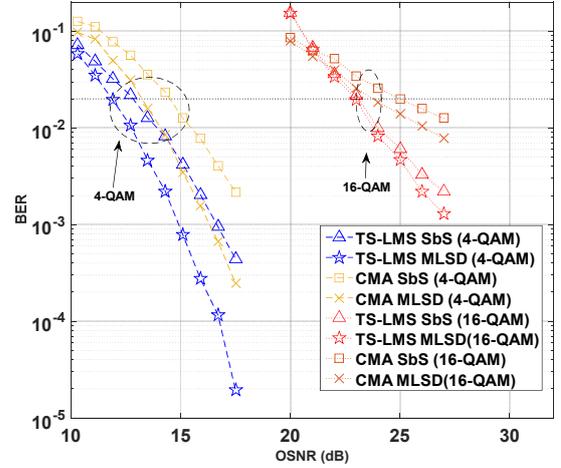

Fig. 4. BER vs. OSNR for QDB FTN systems using the CMA with differential coding and TS-LMS algorithms, respectively.

The convergence performance of the TS-LMS algorithm is compared with the standalone DD-LMS algorithm under different polarization alignment scenarios. Fig. 3 depicts the probability of convergence as a function of the input signal's SOP offset from the principal axis of the DGD element (Δϕ) using both algorithms. 1000 trials are performed for each value of Δϕ to obtain the results. As we can see from the figure, the TS-LMS algorithm always converges successfully, while the probability of convergence when using the standalone DD-LMS is less than 0.25% at an offset from 25 to 65 degrees. Moreover, the system performance comparison between the TS-LMS algorithm and the CMA with differential coding is conducted in simulation. To evaluate both systems when there are cycle slips, the laser linewidth is increased to 1 MHz to introduce higher phase noise. Fig. 4 shows the BER as a function of the OSNR for QDB FTN 4-QAM and 16-QAM systems using both algorithms. In simulations, cycle slips are observed when the blind phase estimation algorithms are used with the CMA, and the error bursts caused by the cycle slips are avoided by differential coding. On the other hand, no cycle slip is observed when the TS-LMS algorithm and the pilot-assisted DA ML algorithmare used. From the results, we can see that the required OSNR at a BER of $2\times10^{-2}$ using the TS-LMS algorithm is 1.7 dB and 1.3 dB lower than using the CMA with differential coding for 4-QAM systems with SbS and MLSD, respectively. For 16-QAM systems, the BER performance is improved by 1.8 dB and 0.8 dB with SbS and MLSD, respectively.

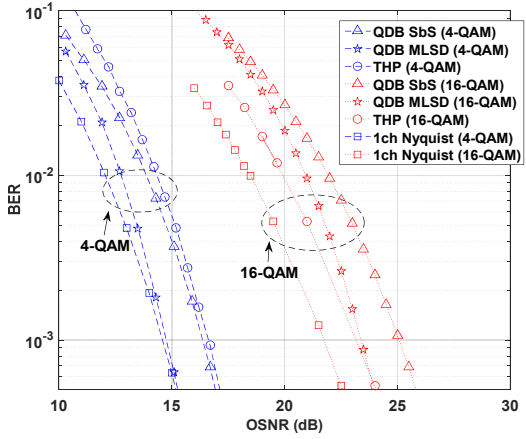

Fig. 5. BER vs. OSNR of FTN systems with QDB and THP.

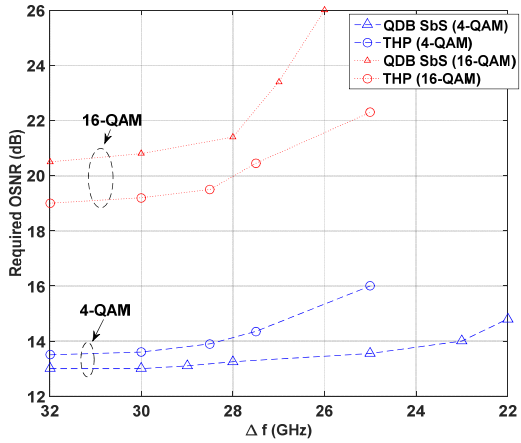

Fig. 6. Required OSNR for a BER of $2\times10^{-2}$ as a function of channel.

In Fig. 5 we illustrate the system performance of the QDB and THP systems. For 4-QAM,QDB with either SbS or MLSD outperforms THP by 0.5 dB and 2.4 dB at a BER of $2\times10^{-2}$, respectively. On the other hand, the performance of QDB, even with MLSD,is 1 dB worse than THP at a BER of $2\times10^{-2}$ for 16-QAM. An explanation is related to the fact that the THP operation maintains the symmetric property of the Gray mapping of 16-QAM, whereas the QDB operation does not. Additionally, we investigate the relationship between the required OSNR for a BER of $2\times10^{-2}$ and $\Delta f$, as shown in Fig. 6. For 4-QAM with QDB, $\Delta f$ can be reduced to 23 GHz at anOSNR penalty of 1 dB with respect to a Nyquist system. $\Delta f$ can be compressed to 28 GHzat 1 dB OSNR penalty for 16-QAM QDB systems. On the other hand, THP allows $\Delta f$ to be 27.5 GHz and 28 GHz for 4-QAM and 16-QAM at an OSNR penalty of 1 dB, respectively.

## IV. CONCLUSIONS

A training symbol-based equalization algorithm has been derived for polarization de-multiplexing in QDB PMD-FTN systems. Comparison with the standalone DD-LMS algorithm shows that the proposed TS-LMS algorithm converges well under differentpolarization alignment conditions. The required OSNR is decreased by about 1.7dB and 1.3 dB using the TS-LMS algorithm for 4-QAM systems, and 1.8 dB and 0.8 dB for 16-QAM systems with SbS and MLSD, respectively, compared to the conventional CMA with differential coding.Further investigation on the comparison with THP shows that QDB is preferable for 4-QAM systems as it provides an improved BER performance, whereas the opposite holds for 16-QAM systems.